\documentclass{PoS}
\usepackage{slashed}
\usepackage{epsfig}
\usepackage{amssymb}
\usepackage{fontenc}
\usepackage{times}
\usepackage{mathptmx}
\usepackage{graphicx}

\title{Baryon chiral perturbation theory}

\ShortTitle{Baryon chiral perturbation theory}

\author{\speaker{Stefan Scherer}%
         \\
         Institut f\"ur Kernphysik, Johannes
         Gutenberg-Universit\"at, D-55099 Mainz, Germany
        \\
        E-mail: \email{scherer@kph.uni-mainz.de}}

\abstract{
   We provide an introduction to the power-counting issue in baryon chiral perturbation
theory and discuss some recent developments in the manifestly Lorentz-invariant
formulation of the one-nucleon sector.
   As explicit applications we consider the chiral expansion of the nucleon mass,
its convergence properties, and calculations of the electromagnetic and
axial form factors of the nucleon.
}

\FullConference{6th International Workshop on Chiral Dynamics\\
                 July 6-10 2009\\
                 Bern, Switzerland}

\begin{document}

\section{Introduction}
   Effective field theory (EFT) is a powerful tool in the description of the
strong interactions at low energies.
   The central idea is due to Weinberg \cite{Weinberg:1978kz}:
   \begin{quote}
"...  if one writes down the most general possible Lagrangian,
including all terms consistent with assumed symmetry principles, and
then calculates matrix elements with this Lagrangian to any given
order of perturbation theory, the result will simply be the most
general possible S--matrix consistent with analyticity, perturbative
unitarity, cluster decomposition and the assumed symmetry
principles."
\end{quote}
    The prerequisite for an effective field theory program
is (a) a knowledge of the most general effective Lagrangian and (b)
an expansion scheme for observables in terms of a consistent power
counting method.
   The application of these ideas to the interactions among the Goldstone bosons
   of spontaneous chiral symmetry breaking in QCD results in
mesonic chiral perturbation theory (ChPT)
\cite{Weinberg:1978kz,Gasser:1983yg} (see, e.g.,
Refs.~\cite{Scherer:2002tk,Bijnens:2006zp,Scherer:2009bt}
for an introduction and overview).
   The combination of dimensional regularization with the modified
minimal subtraction scheme of ChPT \cite{Gasser:1983yg} leads to
a straightforward correspondence between the loop expansion and the
chiral expansion in terms of momenta and quark masses at a fixed ratio,
and thus provides a consistent power counting for renormalized quantities.

   The situation gets more complicated once other hadronic degrees of freedom beyond
the Goldstone bosons are considered.
   Together with such hadrons, another scale of the order of the chiral symmetry
breaking scale $\Lambda_\chi$ enters the problem and the methods of
the pure Goldstone-boson sector cannot be transferred one to one.
   For example, in the extension to the one-nucleon sector
the correspondence between the loop expansion and the chiral
expansion, at first sight, seems to be lost: higher-loop diagrams
can contribute to terms as low as ${\cal O}(q^2)$
\cite{Gasser:1987rb}.
   For a long time this was interpreted as the absence of a systematic power
counting in the relativistic formulation of ChPT.
   However, over the last decade new developments in devising a suitable renormalization
scheme have led to a simple and consistent power counting for the
renormalized diagrams of a manifestly Lorentz-invariant approach.

\section{Renormalization and power counting}

  The effective Lagrangian relevant to the one-nucleon sector
consists of the sum of the purely mesonic and $\pi N$ Lagrangians,
respectively,
\begin{equation}
\label{4:2:full_lagrangian} {\cal L}_{\rm eff} ={\cal L}_{\pi}+{\cal
L}_{\pi N} ={\cal L}_\pi^{(2)}+{\cal L}_\pi^{(4)}+\cdots +{\cal
L}_{\pi N}^{(1)}+{\cal L}_{\pi N}^{(2)} +\cdots,
\end{equation}
which are organized in a derivative and quark-mass expansion.
   Tree-level calculations involving the sum ${\cal L}_\pi^{(2)}+{\cal L}_{\pi N}^{(1)}$
reproduce the current algebra results.
   When studying higher orders in perturbation theory in terms of loop corrections one encounters
ultraviolet divergences.
   In the process of renormalization the
counter terms are adjusted such that they absorb all the ultraviolet
divergences occurring in the calculation of loop diagrams.
   This will be possible, because the Lagrangian includes all
of the infinite number of interactions allowed by symmetries
\cite{Weinberg:1995mt}.
   Moreover, when renormalizing, we still have the freedom of choosing
a renormalization condition.
   The power counting is intimately connected with choosing a suitable
renormalization condition.

\subsection{\label{genct}The generation of counter terms}
   Let us briefly recall the renormalization procedure
in terms of the lowest-order $\pi N$ Lagrangian
${\cal L}_{\pi N}^{(1)}$.
   At the beginning, the (total effective) Lagrangian is formulated in terms of bare
(i.e.~unrenormalized) parameters and fields.
   After expressing the bare
parameters and bare fields in terms of renormalized quantities,
the Lagrangian decomposes into the sum of basic and counter-term
Lagrangians (see,
e.g., Refs.\ \cite{Weinberg:1995mt}, \cite{Collins:xc} for details).
   For example, the basic Lagrangian of lowest order reads
\begin{eqnarray}
\label{4:2:1:lbasic} {\cal L}_{\pi N\, \rm basic}^{(1)}&=& \bar \Psi \left(
i\gamma^\mu\partial_\mu - m -\frac{1}{2}
\frac{\texttt{g}_{A}}{F}\gamma^\mu\gamma_5\partial_\mu
\phi_i\tau_i\right) \Psi+\cdots,
\end{eqnarray}
where the ellipsis refers to terms containing external fields and
higher powers of the pion fields.
   We choose the renormalization condition such that
$m$, $\texttt{g}_A$, and $F$ denote the chiral limit of the physical
nucleon mass, the axial-vector coupling constant, and the pion-decay
constant, respectively.
   Expanding the counter-term Lagrangian
in powers of the renormalized coupling constants generates an
infinite series.
   By adjusting the expansion coefficients suitably, the individual
terms are responsible for the subtractions of loop diagrams.

\subsection{Power counting for renormalized diagrams}
\label{subsection_pcrd}
   Whenever we speak of renormalized diagrams, we refer to
diagrams which have been calculated with a basic Lagrangian and to
which the contribution of the counter-term Lagrangian has been
added.
   Counter-term contributions are typically denoted by a cross.
   One also says that the diagram has been subtracted, i.e., the unwanted
contribution has been removed with the understanding that this can
be achieved by a suitable choice for the coefficient of the
counter-term Lagrangian.
   In this context the {\em finite} pieces of the renormalized
couplings are adjusted such that the renormalized diagrams satisfy
the following power counting:
   a loop integration in $n$ dimensions counts as $q^n$,
pion and nucleon propagators count as $q^{-2}$ and $q^{-1}$,
respectively, vertices derived from ${\cal L}_{\pi}^{(2k)}$ and
${\cal L}_{\pi N}^{(k)}$ count as $q^{2k}$ and $q^k$, respectively.
   Here, $q$ collectively stands for a small quantity such as the pion
   mass, small external four-momenta of the pion, and small external
three-momenta of the nucleon.
   The power counting does not uniquely fix the renormalization scheme,
i.e., there are different renormalization schemes leading to the
above specified power counting.

\subsection{The power-counting problem}
\begin{figure}[t]
\begin{center}
\epsfig{file=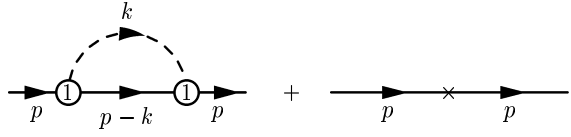,width=0.5\textwidth}
\caption{\label{4:2:3:ren_diag}
Renormalized one-loop self-energy diagram. The
number 1 in the interaction blobs refers to ${\cal L}_{\pi
N}^{(1)}$.}
\end{center}
\end{figure}

   In the mesonic sector, the combination of dimensional regularization and
the modified minimal subtraction scheme $\widetilde{\mbox{MS}}$
leads to a straightforward correspondence between the chiral and loop expansions.
   By discussing the one-loop contribution of Fig.~\ref{4:2:3:ren_diag}
to the nucleon self energy, we will see that this correspondence, at first
sight, seems to be lost in the baryonic sector.
   According to the power counting specified above, after renormalization,
we would like to have the order $D=n\cdot 1-2\cdot 1-1\cdot 1+1\cdot 2=n-1.$
   An explicit calculation yields
\begin{displaymath}
\Sigma_{\rm loop}= -\frac{3 \texttt{g}_{A}^2}{4 F^2}\left\{
(\slashed{p}+m)I_N+M^2(\slashed{p}+m)I_{N\pi}-\frac{(p^2-m^2)\slashed{p}}{2p^2}[
(p^2-m^2+M^2)I_{N\pi}+I_N-I_\pi]\right\},
\end{displaymath}
where $M^2=2B\hat m$ is the lowest-order expression for the squared
pion mass in terms of the low-energy coupling constant $B$ and the
average light-quark mass $\hat m$ \cite{Gasser:1983yg}.
   The relevant loop integrals are defined as
\begin{eqnarray}
\label{Ipi}
I_\pi&=& \mu^{4-n}\int\frac{\mbox{d}^nk}{(2\pi)^n}\frac{i}{k^2-M^2+i0^+},\\
\label{IN}
I_N&=& \mu^{4-n}\int\frac{\mbox{d}^nk}{(2\pi)^n}\frac{i}{k^2-m^2+i0^+},\\
\label{INpi} I_{N\pi}&=&\mu^{4-n}\int\frac{\mbox{d}^nk}{(2\pi)^n}
\frac{i}{[(k-p)^2-m^2+i0^+]}\frac{1}{k^2-M^2+i0^+}.
\end{eqnarray}
   The application of the $\widetilde{\rm MS}$ renormalization scheme of ChPT
\cite{Gasser:1983yg,Gasser:1987rb}---indicated by ``r''---yields
\begin{displaymath}
\Sigma_{\rm loop}^r=-\frac{3 \texttt{g}_{Ar}^2}{4 F^2}
\left[M^2(\slashed{p}+m)I_{N\pi}^r+\cdots\right].
\end{displaymath}
   The expansion of $I_{N\pi}^r$ is given by
\begin{displaymath}
\label{4:2:3:INpExp}
    I_{N\pi}^r=\frac{1}{16\pi^2}\left(-1+\frac{\pi M}{m}+\cdots\right),
\end{displaymath}
resulting in $\Sigma_{\rm loop}^r={\cal O}(q^2)$.
   In other words, the $\widetilde{\rm MS}$-renormalized result does not
produce the desired low-energy behavior which, for a long time, was
interpreted as the absence of a systematic power counting in the
relativistic formulation of ChPT.

   The expression for the nucleon mass $m_N$ is obtained by solving
the equation
\begin{displaymath}
\label{4:2:3:MassDef}
m_N-m-\Sigma(m_N)=0,
\end{displaymath}
from which we obtain for the nucleon mass in the $\widetilde{\rm
MS}$ scheme \cite{Gasser:1987rb},
\begin{equation}
\label{4:2:3:MassMStilde}
    m_N=m-4c_{1r}M^2+
    \frac{3\texttt{g}_{Ar}^2M^2}{32\pi^2F^2}m
    -\frac{3\texttt{g}_{Ar}^2M^3}{32\pi^2F^2}.
\end{equation}
   At ${\cal O}(q^2)$, Eq.~(\ref{4:2:3:MassMStilde}) contains besides the undesired loop
contribution proportional to $M^2$ the tree-level contribution $-4c_{1r}M^2$
from the next-to-leading-order Lagrangian ${\cal L}_{\pi N}^{(2)}$.

   The solution to the power-counting problem is the observation
that the term violating the power counting, namely, the third on the
right-hand side of Eq.~(\ref{4:2:3:MassMStilde}), is \emph{analytic}
in the quark mass and can thus be absorbed in counter terms.
   In addition to the $\widetilde{\rm MS}$ scheme we have to perform an additional
{\em finite} renormalization.
   For that purpose we rewrite
\begin{equation}
\label{4:2:3:cRenorm}
    c_{1r}=c_1+\delta c_1,\quad \delta c_1 =\frac{3 m {\texttt g}_A^2}{128 \pi^2 F^2}+\cdots
\end{equation}
in Eq.~(\ref{4:2:3:MassMStilde}) which then gives the final result
for the nucleon mass at ${\cal O}(q^3)$:
\begin{equation}
\label{4:2:3:MassFinal}
    m_N=m-4c_{1}M^2
    -\frac{3\texttt{g}_{A}^2M^3}{32\pi^2F^2}.
\end{equation}
   We have thus seen that the validity of a power-counting scheme is intimately
connected with a suitable renormalization condition.
   In the case of the nucleon mass, the $\widetilde{\rm MS}$ scheme alone does not
suffice to bring about a consistent power counting.

\subsection{Solutions to the power-counting problem}

\subsubsection{Heavy-baryon approach}
\label{hb}
   The first solution to the power-counting problem
was provided by the heavy-baryon formulation of ChPT
\cite{Jenkins:1990jv,Bernard:1992qa}.
   The basic idea consists in dividing an external nucleon four-momentum
into a large piece close to on-shell kinematics and a soft residual
contribution: $p = m v +k_p$, $v^2=1$, $v^0\ge 1$ [often $v^\mu =
(1,0,0,0)$].
   The relativistic nucleon field is expressed in terms of
velocity-dependent fields,
\begin{displaymath}
\Psi(x)=e^{-im v \cdot x} ({\cal N}_v +{\cal H}_v),
\end{displaymath}
with
\begin{displaymath}
{\cal N}_v=e^{+im v\cdot x}\frac{1}{2}(1+v\hspace{-.5em}/)\Psi,\quad
{\cal H}_v=e^{+im v\cdot x}\frac{1}{2}(1-v\hspace{-.5em}/)\Psi.
\end{displaymath}
   Using the equation of motion for ${\cal H}_v$, one can
eliminate ${\cal H}_v$ and obtain a Lagrangian for ${\cal N}_v$
which, to lowest order, reads \cite{Bernard:1992qa}
\begin{displaymath}
\widehat{\cal L}^{(1)}_{\pi N}=\bar{\cal N}_v(iv\cdot D +
\texttt{g}_A S_v\cdot u) {\cal N}_v+{\cal O}(1/m),\quad
S^\mu_v=\frac{i}{2}\gamma_5\sigma^{\mu\nu}v_\nu.
\end{displaymath}
   The result of the heavy-baryon reduction is a $1/m$ expansion of the
Lagrangian similar to a Foldy-Wouthuysen expansion.
   In higher orders in the chiral expansion, the expressions due to
$1/m$ corrections of the Lagrangian become increasingly complicated
\cite{Fettes:2000gb}.
   Moreover---and what is more important---the approach sometimes generates problems
regarding analyticity \cite{Becher:1999he}.

\subsubsection{Master integral}
\label{subsubsection_master_integral}
   We have seen that the modified minimal subtraction scheme $\widetilde{\rm MS}$
does not produce the desired power counting.
   We will discuss the power-counting problem in terms of the dimensionally regularized
one-loop integral
\begin{eqnarray}
\label{4:3:2:Hdef} H(p^2,m^2,M^2;n) &\equiv&
-i\int \frac{\mbox{d}^n k}{(2\pi)^n} \frac{1}{[k^2-2p\cdot k
+(p^2-m^2)+i0^+](k^2-M^2+i0^+)}.
\end{eqnarray}
   We are interested in nucleon four-momenta close to the mass-shell condition,
$p^2\approx m^2$, counting $p^2-m^2$ as ${\cal O}(q)$ and $M^2$ as
${\cal O}(q^2)$.
   Making use of the Feynman parametrization
\begin{displaymath}
{1\over ab}=\int_0^1 {\mbox{d}z\over [az+b(1-z)]^2}
\end{displaymath}
with $a=k^2-2p\cdot k+(p^2-m^2)+i0^+$ and $b=k^2-M^2+i0^+$, interchanging the
order of integrations, and performing the shift $k\to k+zp$,
one obtains
\begin{equation}
\label{4:3:2:HPmMn} H(p^2,m^2,M^2;n)=\frac{1}{(4\pi)^{\frac{n}{2}}}
\Gamma\left(2-\frac{n}{2}\right) \int_0^1 \mbox{d}z
[A(z)-i0^+]^{\frac{n}{2}-2},
\end{equation}
where $A(z)=z^2 p^2-z(p^2-m^2+M^2)+M^2$.
   The relevant properties can nicely be displayed at the threshold
$p^2_{\rm thr}=(m+M)^2$, where $A(z)=[z(m+M)-M]^2$ is particularly
simple.
    Splitting the integration interval into $[0,z_0]$ and $[z_0,1]$ with
$z_0=M/(m+M)$, we have, for $n>3$,
\begin{eqnarray*}
\int_0^1 \mbox{d}z [A(z)]^{\frac{n}{2}-2}&=&\int_0^{z_0}\mbox{d}z
[M-z(m+M)]^{n-4}
+\int_{z_0}^1\mbox{d}z [z(m+M)-M]^{n-4}\\
&=&\frac{1}{(n-3)(m+M)}(M^{n-3}+m^{n-3}),
\end{eqnarray*}
yielding, through analytic continuation, for arbitrary $n$
\begin{equation}
\label{4:3:2:defhthr}
 H((m+M)^2,m^2,M^2;n)=
\frac{\Gamma\left(2-\frac{n}{2}\right)}{(4\pi)^{\frac{n}{2}}(n-3)}
\left(\frac{M^{n-3}}{m+M}+\frac{m^{n-3}}{m+M}\right).
\end{equation}
   The first term, proportional to $M^{n-3}$, is defined as the so-called
infrared singular part $I$.
   Since $M\to 0$ implies $p^2_{\rm thr} \to m^2$ this term is
singular for $n\leq 3$.
   The second term, proportional to $m^{n-3}$, is defined as the
infrared regular part $R$.

\subsubsection{Infrared regularization}
   The {\em formal} definition of Becher and Leutwyler \cite{Becher:1999he}
for the infrared singular and regular parts for arbitrary $p^2$
makes use of the Feynman parametrization of Eq.\
(\ref{4:3:2:HPmMn}).
   The resulting integral over the Feynman parameter $z$ is rewritten as
\begin{equation}
\label{HIR}
H=\int_0^1 \mbox{d}z \cdots = \int_0^\infty \mbox{d}z \cdots
- \int_1^\infty \mbox{d}z \cdots\equiv I+R.
\end{equation}
   What distinguishes $I$ from $R$ is that, for non-integer values of
$n$, the chiral expansion of $I$ gives rise to non-integer powers of
${\cal O}(q)$, whereas the regular part $R$ may be expanded in an
ordinary Taylor series.
   For the threshold integral, this can nicely be seen by expanding
$I_{\rm thr}$ and $R_{\rm thr}$ in the pion mass counting as ${\cal
O}(q)$.
   On the other hand, it is the regular part which does not satisfy
the counting rules.
    The basic idea of the infrared renormalization consists of replacing
the general integral $H$ of Eq.\ (\ref{4:3:2:HPmMn}) by its infrared
singular part $I$ and dropping
the regular part $R$.
   In the low-energy region $H$ and $I$ have the same analytic properties
whereas the contribution of $R$, which is of the type of an infinite
series in the momenta, can be included by adjusting the coefficients
of the most general effective Lagrangian.
   This is the infrared renormalization condition.
   As discussed in detail in Ref.\ \cite{Becher:1999he}, the method can
be generalized to an arbitrary one-loop graph.

\subsubsection{Extended on-mass-shell scheme}
   In the following, we will concentrate on yet another solution which
has been motivated in Ref.\ \cite{Gegelia:1999gf} and has been worked out in
detail in Ref.\ \cite{Fuchs:2003qc}.
   The central idea consists of performing additional subtractions beyond
the $\widetilde{\rm MS}$ scheme such that the renormalized diagrams satisfy
the power counting (``{\em choosing a suitable renormalization condition}'').
   Terms violating the power counting are analytic in small
quantities and can thus be absorbed in a renormalization of
counter terms.
   In order to illustrate the approach, let us consider as an example
the integral of Eq.~(\ref{4:3:2:Hdef}) in the chiral limit,
\begin{displaymath}
H(p^2,m^2,0;n)= \int \frac{\mbox{d}^n k}{(2\pi)^n}
\frac{i}{[k^2-2p\cdot k +(p^2-m^2)+i0^+][k^2+i0^+]},
\end{displaymath}
where
\begin{displaymath}
\Delta=\frac{p^2-m^2}{m^2}={\cal O}(q)
\end{displaymath}
is a small quantity.
   We want the (renormalized) integral to be of order
$D=n-1-2=n-3$.
   The result of the integration is of the form (see Ref.\
\cite{Fuchs:2003qc} for details)
$H\sim F(n,\Delta)+\Delta^{n-3}G(n,\Delta)$,
where $F$ and $G$ are hypergeometric functions and are analytic in $\Delta$ for
any $n$.
   Hence, the part containing $G$ for noninteger $n$ is proportional to
a noninteger power of $\Delta$ and satisfies the power counting.
   The observation central for the setting up of a systematic method is the fact
that the part proportional to $F$ can be obtained by first expanding
the integrand in small quantities and {\em then} performing the
integration for each term \cite{Gegelia:1994zz}.
   It is this part which violates the power counting, but, since
it is analytic in $\Delta$, the power-counting violating pieces can
be absorbed in the counter terms.
   This observation suggests the following procedure: expand the integrand in
small quantities and subtract those (integrated) terms whose order is
smaller than suggested by the power counting.
   Since the subtraction point is $p^2=m^2$, the renormalization condition
is denoted ``extended on-mass-shell'' (EOMS) scheme in analogy with
the on-mass-shell renormalization scheme in renormalizable theories.
   In the present case, the subtraction term reads
\begin{displaymath}
H^{\rm subtr}=\int \frac{\mbox{d}^n k}{(2\pi)^n}\left.
\frac{i}{[k^2-2p\cdot k +i0^+][k^2+i0^+]}\right|_{p^2=m^2}
\end{displaymath}
and the renormalized integral is written as
\begin{displaymath}
H^R=H-H^{\rm subtr}={\cal O}(q^{n-3}).
\end{displaymath}

\subsection{Remarks}

Using a suitable renormalization condition one
obtains a consistent power counting in manifestly Lorentz-invariant
baryon ChPT including, e.g., (axial) vector mesons \cite{Fuchs:2003sh} or
the $\Delta(1232)$ resonance \cite{Hacker:2005fh} as explicit
degrees of freedom. The infrared regularization of Becher and Leutwyler
\cite{Becher:1999he} has been reformulated in a form analogous to
the EOMS renormalization \cite{Schindler:2003xv}.
The application of both
infrared and extended on-mass-shell renormalization schemes to
multi-loop diagrams was explicitly demonstrated by means of a
two-loop self-energy diagram \cite{Schindler:2003je}.
A treatment of unstable particles such as the rho meson is possible in terms
of the complex-mass renormalization \cite{Djukanovic:2009zn}.

\section{Applications}
\label{section_applications}
   In the following we will illustrate a few selected applications
of the manifestly Lorentz-invariant framework to the one-nucleon
sector.

\subsection{Nucleon mass at ${\cal O}(q^4)$}
\begin{figure}[t]
\begin{center}
\epsfig{file=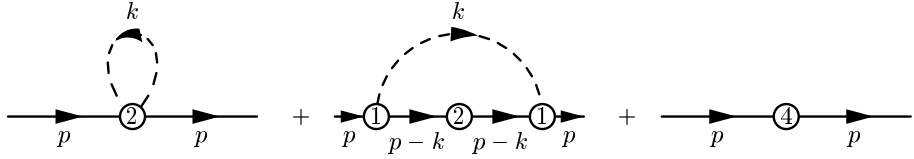,width=0.8\textwidth}
\caption{\label{5:1:1:SEDiagrams} Contributions to the nucleon self
energy at ${\cal O}(q^4)$. The number $n$ in the interaction blobs
refers to ${\cal L}_{\pi N}^{(n)}$. The Lagrangian ${\cal L}_{\pi
N}^{(2)}$ does not produce a contribution to the $\pi NN$ vertex.}
\end{center}
\end{figure}

   A full one-loop calculation of the nucleon mass also includes
${\cal O}(q^4)$ terms (see Fig.~\ref{5:1:1:SEDiagrams}).
   The quark-mass expansion up to and including ${\cal O}(q^4)$ is given by
\begin{equation}
\label{mnoq4} m_N=m+k_1 M^2+k_2 M^3+k_3
M^4\ln\left(\frac{M}{m}\right)+k_4 M^4 +{\cal O}(M^5),
\end{equation}
where the coefficients $k_i$ in the EOMS scheme read
\cite{Fuchs:2003qc}
\begin{eqnarray}
\label{parki} k_1&=&-4 c_1,\quad k_2=-\frac{3
{\texttt{g}_A}^2}{32\pi F^2},\quad k_3=-\frac{3}{32\pi^2 F^2
m}\left(\texttt{g}_A^2-8c_1m+c_2m+4 c_3m\right),
\nonumber\\
k_4&=&\frac{3 {\texttt{g}_A}^2}{32 \pi^2 F^2 m}(1+4 c_1 m)
+\frac{3}{128\pi^2F^2}c_2-\hat{e}_1.
\end{eqnarray}
   Here, $\hat{e}_1= 16 e_{38}+2e_{115}+2e_{116}$ is a linear combination
of ${\cal O}(q^4)$ coefficients \cite{Fettes:2000gb}.
   A comparison with the results using the infrared regularization \cite{Becher:1999he}
shows that the lowest-order correction ($k_1$ term) and those terms
which are non-analytic in the quark mass $\hat{m}$ ($k_2$ and $k_3$
terms) coincide.
   On the other hand, the analytic $k_4$ term ($\sim M^4$) is different.
   This is not surprising; although both renormalization schemes satisfy
the power counting specified in Sec.~\ref{subsection_pcrd}, the use
of different renormalization conditions is compensated by different
values of the renormalized parameters.

   For an estimate of the various contributions of Eq.\ (\ref{mnoq4}) to the nucleon mass,
we make use of the parameter set
\begin{equation}
\label{parametersci} c_1=-0.9\,m_N^{-1},\quad c_2=2.5\,
m_N^{-1},\quad c_3=-4.2\, m_N^{-1},\quad c_4=2.3\, m_N^{-1},
\end{equation}
which was obtained in Ref.\ \cite{Becher:2001hv} from a (tree-level)
fit to the $\pi N$ scattering threshold parameters.
   Using the numerical values
\begin{equation}
\label{numericalvalues} g_A=1.267,\quad F_\pi=92.4\,\mbox{MeV},\quad
m_N=m_p=938.3\,\mbox{MeV},\quad M_\pi=M_{\pi^+}=139.6\,\mbox{MeV},
\end{equation}
one obtains for the mass of nucleon in the chiral limit (at fixed
$m_s\neq 0$) \cite{Fuchs:2003kq}:
\begin{equation}
m=m_N-\Delta m =[938.3-74.8+15.3+4.7+1.6-2.3\pm 4]\,\mbox{MeV}
=(883\pm 4)\, \mbox{MeV}
\end{equation}
with $\Delta m=(55.5\pm 4)\,\mbox{MeV}$.
   Here, we have made use of an estimate for $\hat{e}_1M^4=(2.3\pm 4)$ MeV obtained from
the sigma term $\sigma=(45\pm 8)$ MeV.
   Note that errors due to higher-order corrections are not taken into account.

\subsection{Chiral expansion of the nucleon mass to ${\cal O}(q^6)$}
    So far, essentially all of the manifestly Lorentz-invariant calculations have been
restricted to the one-loop level.
   One of the exceptions is the chiral expansion of the nucleon mass which,
in the framework of the reformulated infrared regularization, has
been calculated up to and including ${\cal O}(q^6)$
\cite{Schindler:2006ha,Schindler:2007dr}:
\begin{equation}
\label{Mass:Exp}
    m_N = m +k_1 M^2 +k_2 \,M^3 +k_3 M^4 \ln\frac{M}{\mu}
+ k_4 M^4  + k_5 M^5\ln\frac{M}{\mu} + k_6 M^5 + k_7 M^6
\ln^2\frac{M}{\mu}+ k_8 M^6 \ln\frac{M}{\mu} + k_9 M^6.
\end{equation}
   We refrain from displaying the lengthy expressions for the coefficients
$k_i$ but rather want to discuss a few general implications
\cite{Schindler:2007dr}.
   Chiral expansions like Eq.~(\ref{Mass:Exp}) currently play an important
role in the extrapolation of lattice QCD results to physical quark
masses.
   Unfortunately, the numerical contributions from higher-order terms cannot be
calculated so far since, starting with $k_4$, most expressions in
Eq.~(\ref{Mass:Exp}) contain unknown low-energy coupling constants
(LECs) from the Lagrangians of ${\cal O}(q^4)$ and higher.
   The coefficient $k_5$ is free of higher-order LECs
and is given in terms of the axial-vector coupling constant
$\texttt{g}_A$ and the pion-decay constant $F$:
\begin{displaymath}
 k_5 = \frac{3 \texttt{g}_A^2}{1024\pi^3 F^4}\,\left(16\texttt{g}_A^2-3\right).
 \end{displaymath}
   While the values for both $\texttt{g}_A$ and $F$ should be taken in the chiral
limit, we evaluate $k_5$ using the physical values $g_A=1.2695(29)$
and $F_\pi=92.42(26)$ MeV.
   Setting $\mu=m_N$, $m_N=(m_p+m_n)/2=938.92$ MeV, and $M=M_{\pi^+}=139.57$ MeV
we obtain $k_5 M^5 \ln(M/m_N) = -4.8$ MeV.
   This amounts to approximately $31$\% of the leading non-analytic contribution
at one-loop order, $k_2 M^3$.
\begin{figure}[t]
\begin{center}
\epsfig{file=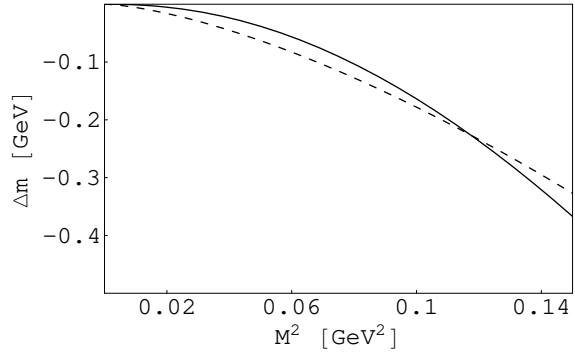,width=0.5\textwidth}
\end{center}
\caption{Pion mass dependence of the term $k_5 M^5 \ln(M/m_N)$
(solid line) for $M<400\,\mbox{MeV}$. For comparison also the term
$k_2 M^3$ (dashed line) is shown.\label{Mass:k2k5Low}}
\end{figure}
   Figure \ref{Mass:k2k5Low} shows the pion mass dependence of the term
$k_5 M^5 \ln(M/m_N)$ (solid line) in comparison with the term $k_2
M^3$ (dashed line) for pion masses below $400\,\mbox{MeV}$ which is
considered a region where chiral extrapolations are valid (see,
e.g., Refs.~\cite{Meissner:2005ba,Djukanovic:2006xc}).
   We see that already at $M \approx 360\,\mbox{MeV}$ the
term $k_5 M^5 \ln(M/m_N)$ becomes as large as the leading
non-analytic term at one-loop order, $k_2 M^3$, indicating the
importance of the fifth-order terms at unphysical pion masses.
   The results for the renormalization-scheme-independent terms agree
with the heavy-baryon ChPT results of Ref.~\cite{McGovern:1998tm}.

\subsection{Probing the convergence of perturbative series}
   The issue of the convergence of perturbative calculations
is presently of great interest in the context of chiral extrapolations of
baryon properties (see, e.g., Refs.~\cite{Leinweber:2003dg,Procura:2003ig,Beane:2004ks}).
   A possibility of exploring the convergence of perturbative series
consists of summing up certain sets of an infinite number of diagrams by solving
integral equations exactly and comparing the solutions with the perturbative
contributions \cite{Djukanovic:2006xc}.
\begin{figure}[t]
\begin{center}
\epsfig{file=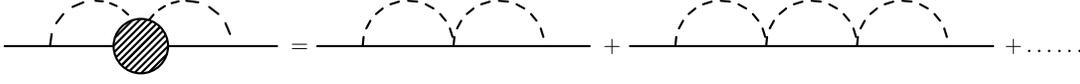, width=\textwidth}
\end{center}
\caption{Iterated contribution to the nucleon self energy.\label{Nse:figdiagrams}}
\end{figure}
   Figure \ref{Nse:figdiagrams} shows a graphical representation
of an iterated contribution to the nucleon self energy originating from the
Weinberg-Tomozawa term in the $\pi N$ scattering amplitude.
   The result is of the form \cite{Djukanovic:2006xc}
\begin{equation}
\delta m = -\frac{3\texttt{g}_{A}^2}{4F^2}
\frac{N}{D}, \label{contributioninmass}
\end{equation}
where $N$ and $D$ are closed expressions in terms of the loop functions of
Eqs.~(\ref{Ipi}) - (\ref{INpi}).
   By expanding Eq.~(\ref{contributioninmass}) in
powers of $1/F^2$ one can identify the contributions of each
diagram separately.
   Using the IR renormalization scheme and substituting $m=883$ MeV,
$m_N=938.3$ MeV, $F=92.4$ MeV, $\texttt{g}_{A}=1.267$ and $M=139.6$ MeV one
obtains
\begin{equation}
\delta m = -0.00233530\,{\rm MeV}=\left( -0.00230219-0.00003305 -
0.00000007 +\cdots\right)\,{\rm MeV}\,. \label{dmBL}
\end{equation}
   The first term in the perturbative expansion reproduces the non-perturbative result
well and the higher-order corrections are clearly suppressed.
   Figure \ref{Nseplot:fig} shows $\delta m$ of
Eq.~(\ref{contributioninmass}) together with the leading contribution (first
diagram in Fig.~\ref{Nse:figdiagrams}) and the leading
non-analytic correction to the nucleon mass $\delta
m_3=-3\,g_A^2\,M^3/(32\,\pi\,F^2)$ \cite{Gasser:1987rb} as
functions of $M$. As can be seen from this figure, up to $M\sim
500$ MeV the non-perturbative sum of higher-order corrections is
suppressed in comparison with the $\delta m_3$ term. Also, the
leading higher-order contribution reproduces the non-perturbative
result quite well. On the other hand, for $M\gtrsim 600$ MeV the
higher-order contributions are no longer suppressed in comparison
with $\delta m_3$.

\begin{figure}
\begin{center}
\epsfig{file=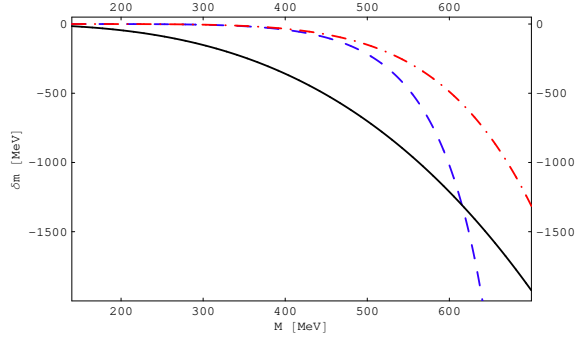, width=0.5\textwidth}
\caption[]{\label{Nseplot:fig} Contributions to the nucleon mass
as functions of $M$. Solid line: ${\cal O}(q^3)$ contribution,
dashed line: $\delta m$ of Eq.~(\ref{contributioninmass});
dashed-dotted line: two-loop diagram of
Fig.~\ref{Nse:figdiagrams}.}
\end{center}
\end{figure}

\subsection{Electromagnetic form factors of the nucleon}
   Imposing the relevant symmetries such as translational invariance,
Lorentz covariance, the discrete symmetries, and current
conservation, the nucleon matrix element of the electromagnetic
current operator ${\cal J}^\mu(x)$ can be parameterized in terms of
two form factors,
\begin{equation}
\label{H1:emff:empar} \langle N(p',s')|{\cal
J}^{\mu}(0)|N(p,s)\rangle= \bar{u}(p',s')\left[F_1^N(Q^2)\gamma^\mu
+i\frac{\sigma^{\mu\nu}q_\nu}{2m_p}F_2^N(Q^2) \right]u(p,s),\quad
N=p,n,
\end{equation}
   where $q=p'-p$, $Q^2=-q^2$, and $m_p$ is the proton mass.
   At $Q^2=0$, the so-called Dirac and Pauli form factors $F_1$ and
$F_2$ reduce to the charge and anomalous magnetic moment in units of
the elementary charge and the nuclear magneton $e/(2m_p)$,
respectively,
\begin{displaymath}
F_1^{p}(0)=1,\quad F_1^{n}(0)=0,\quad F_2^{p}(0)=1.793,\quad
F_2^{n}(0)=-1.913.
\end{displaymath}
   The Sachs form factors $G_E$ and $G_M$ are linear combinations of $F_1$ and
$F_2$,
\begin{displaymath}
G_E^N(Q^2)=F_1^N(Q^2)-\frac{Q^2}{4m_p^2}F_2^N(Q^2),\quad
G_M^N(Q^2)=F_1^N(Q^2)+F_2^N(Q^2), \quad N=p,n,
\end{displaymath}
and, in the non-relativistic limit, their Fourier transforms are
commonly interpreted as the distribution of charge and magnetization
inside the nucleon.

\begin{figure}[t]
\begin{center}
\epsfig{file=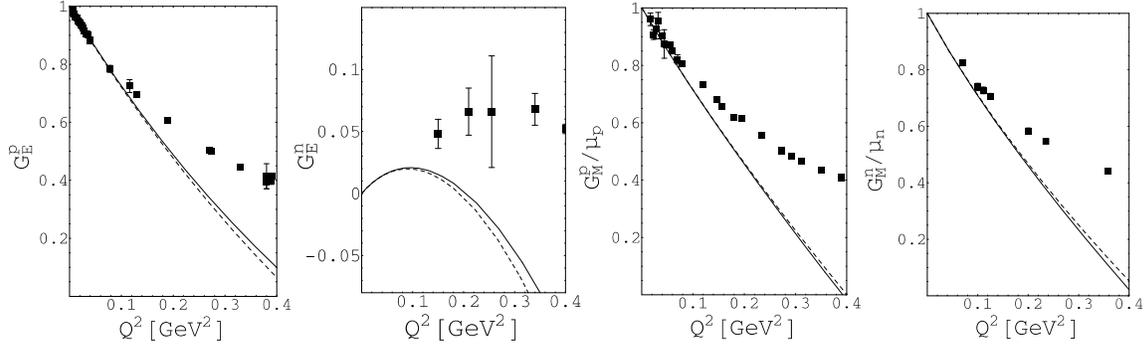,width=\textwidth}
\caption{\label{G_ohne} The Sachs form factors of the nucleon in
manifestly Lorentz-invariant chiral perturbation theory at ${\cal
O}(q^4)$ without vector mesons. Full lines: results in the extended
on-mass-shell scheme; dashed lines: results in infrared
regularization. The experimental data are taken from Ref.\
\cite{Friedrich:2003iz}.}
\end{center}
\end{figure}

   Calculations in Lorentz-invariant baryon ChPT up to
fourth order fail to describe the proton and nucleon form factors
for momentum transfers beyond $Q^2\sim 0.1\, \mbox{GeV}^2$
\cite{Kubis:2000zd,Fuchs:2003ir}.
   Moreover, up to and including ${\cal O}(q^4)$, the most general effective Lagrangian
provides sufficiently many independent parameters such that the
empirical values of the anomalous magnetic moments and the charge
and magnetic radii are fitted rather than predicted.
   Figure \ref{G_ohne} shows the Sachs form factors in the momentum transfer region
$0\,{\rm GeV^2}\le Q^2\le 0.4\,{\rm GeV^2}$ in the EOMS scheme and
the reformulated infrared regularization \cite{Schindler:2005ke}.

\begin{figure}[t]
\begin{center}
\epsfig{file=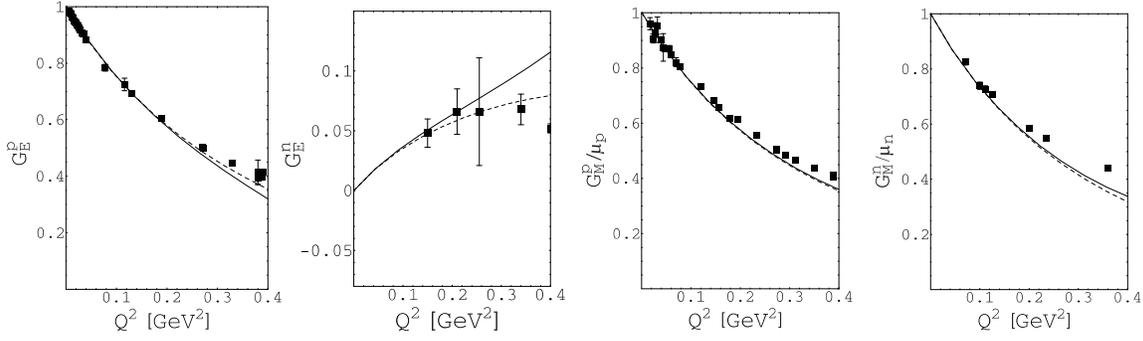,width=\textwidth}
\caption{\label{H1:emff:G:neu} The Sachs form factors of the nucleon
in manifestly Lorentz-invariant chiral perturbation theory at ${\cal
O}(q^4)$ including vector mesons as explicit degrees of freedom.
Full lines: results in the extended on-mass-shell scheme; dashed
lines: results in infrared regularization. The experimental data are
taken from Ref.\ \cite{Friedrich:2003iz}.}
\end{center}
\end{figure}

   In Ref.\ \cite{Kubis:2000zd} it was shown that the
inclusion of vector mesons can result in the re-summation of
important higher-order contributions.
   In Ref.\ \cite{Schindler:2005ke} the electromagnetic form factors of the
nucleon up to fourth order have been calculated in manifestly
Lorentz-invariant ChPT with vector mesons as explicit degrees of
freedom.
   A systematic power counting for the renormalized diagrams has been
implemented using both the extended on-mass-shell renormalization
scheme and the reformulated version of infrared regularization.
   As expected on phenomenological grounds, the quantitative description of the
data has improved considerably for $Q^2\geq 0.1$ GeV$^2$   (see
Fig.~\ref{H1:emff:G:neu}).
   The small difference between the two renormalization schemes is due to the
way how the regular higher-order terms of loop integrals are
treated.
   Numerically, the results are similar to those of Ref.\ \cite{Kubis:2000zd}.
   Due to the renormalization condition, the contribution of the vector-meson
loop diagrams either vanishes (IR) or turns out to be small (EOMS).
   Thus, in hindsight our approach puts the traditional phenomenological
vector-meson-dominance model on a more solid theoretical basis.
   The inclusion of vector-meson degrees of freedom in the present
framework results in a reordering of terms which, in an ordinary
chiral expansion, would show up at higher orders beyond ${\cal
O}(q^4)$.
   It is these terms which change the form factor results favorably for larger
values of $Q^2$.
   It should be noted, however, that this re-organization proceeds according to
well-defined rules so that a controlled, order-by-order, calculation
of corrections is made possible.

\subsection{Axial and induced pseudoscalar form factors}
   Assuming isospin symmetry, the most general
parametrization of the isovector axial-vector current evaluated
between one-nucleon states is given by
\begin{equation}\label{H1_axff_FFDef}
\langle N(p')| A^{\mu,a}(0) |N(p) \rangle = \bar{u}(p')
\left[\gamma^\mu\gamma_5 G_A(Q^2) +\frac{q^\mu}{2m_N}\gamma_5
G_P(Q^2) \right] \frac{\tau^a}{2}u(p),
\end{equation}
where $q=p'-p$, $Q^2=-q^2$, and $m_N$ denotes the nucleon mass.
   $G_A(Q^2)$ is called the axial form factor and
$G_P(Q^2)$ is the induced pseudoscalar form factor.
     The value of the axial form factor at zero momentum transfer is defined as
the axial-vector coupling constant, $g_A=G_A(Q^2=0) =1.2695(29)$,
and is quite precisely determined from neutron beta decay.
   The $Q^2$ dependence of the axial form factor can be obtained
either through neutrino scattering or pion electroproduction
(see, e.g.,  Ref.\ \cite{Bernard:2001rs}).
   The induced pseudoscalar form factor $G_P(Q^2)$ has been investigated in
ordinary and radiative muon capture as well as pion
electroproduction (see Ref.\ \cite{Gorringe:2002xx} for a review).

\begin{figure}[tb]
\begin{minipage}[b]{0.3\textwidth}
\includegraphics[width=\textwidth]{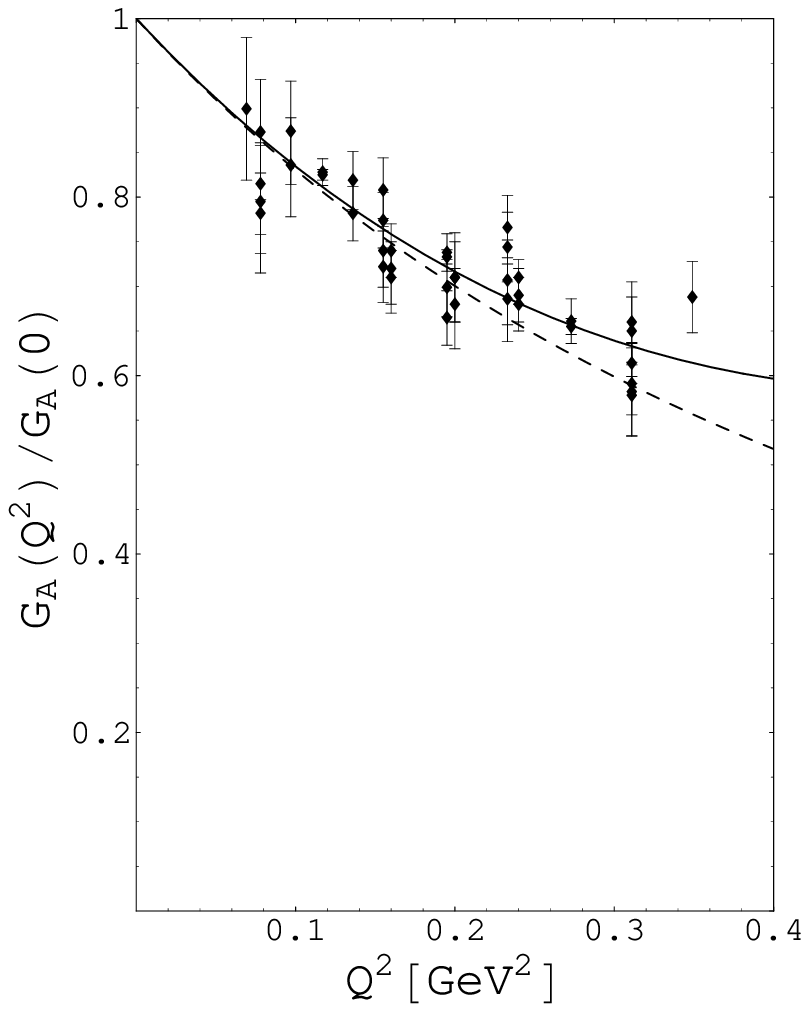}
\end{minipage}
\hspace{2em}
\begin{minipage}[b]{0.5\textwidth}
\includegraphics[width=\textwidth]{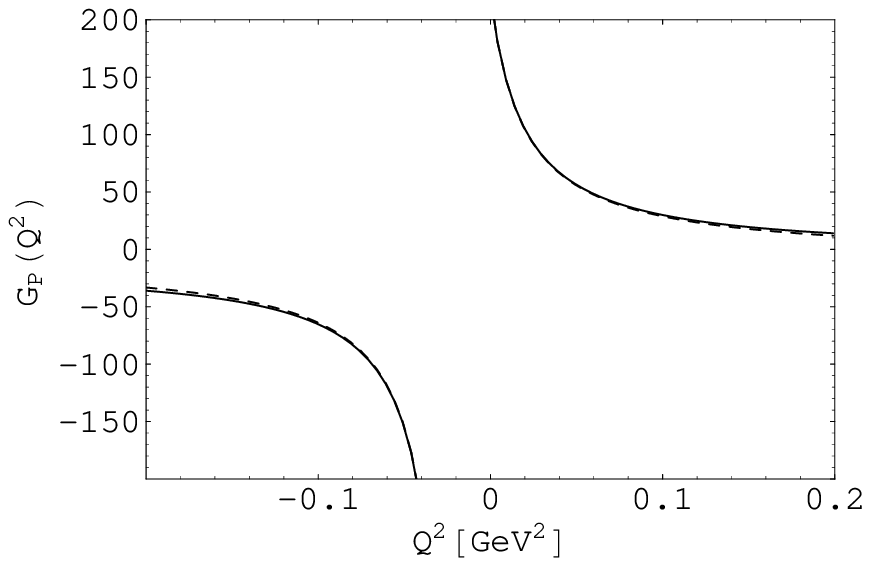}
\end{minipage}
\caption{\label{H1_axff_GAwith}Left panel: Axial form factor $G_A$
in manifestly Lorentz-invariant ChPT at ${\cal O}(q^4)$ including
the axial-vector meson $a_1(1260)$ explicitly. Full line: result in
infrared renormalization, dashed line: dipole parametrization. The
experimental data are taken from Ref.\ \cite{Bernard:2001rs}.
 Right
panel:
   The induced pseudoscalar form factor $G_P$ in manifestly Lorentz-invariant ChPT
at ${\cal O}(q^4)$ including the axial-vector meson $a_1(1260)$
explicitly. Full line: result with axial-vector meson; dashed line:
result without axial-vector meson.
   One can clearly see the dominant pion pole contribution at $Q^2\approx
   -M_\pi^2$.}
\end{figure}

   In Ref.\ \cite{Schindler:2006it} the form factors $G_A$ and
$G_P$ have been calculated in manifestly Lorentz-invariant baryon
ChPT up to and including order ${\cal O}(q^4)$.
   In addition to the standard treatment including the
nucleon and pions, the axial-vector meson $a_1(1260)$ has also been
considered as an explicit degree of freedom.
   The inclusion of the axial-vector meson effectively results in one additional
low-energy coupling constant which has been determined by a fit to
the data for $G_A(Q^2)$.
   The inclusion of the axial-vector meson results in an improved
description of the experimental data for $G_A$ (see
Fig.~\ref{H1_axff_GAwith}), while the contribution to $G_P$ is
small.

\section{Summary and conclusion}
   In the baryonic sector new renormalization conditions have reconciled
the manifestly Lorentz-invariant approach with the standard power
counting.
   We have discussed some results of a two-loop calculation of
the nucleon mass.
   The inclusion of vector and axial-vector mesons as explicit
degrees of freedom leads to an improved phenomenological description
of the electromagnetic and axial form factors, respectively.
   Work on the application to electromagnetic processes such as Compton
scattering and pion production is in progress.

   I would like to thank D.~Djukanovic, T.~Fuchs, J.~Gegelia,
 G.~Japaridze, and M.~R.~Schindler for the
fruitful collaboration on the topics of this talk. This work was
made possible by the financial support from the Deutsche
Forschungsgemeinschaft (SFB 443 and SCHE 459/2-1) and the EU
Integrated Infrastructure Initiative Hadron Physics Project
(contract number RII3-CT-2004-506078).

\end{document}